\documentclass[pss]{wiley2sp}
\usepackage{amsmath}

\usepackage{amsmath,amsfonts,amssymb}

\usepackage{bm}
\usepackage{dsfont}
\usepackage{graphicx}
\usepackage{caption}
\usepackage{xcolor}
\usepackage{braket}
\title{Mie Scattering in the Macroscopic Response and the Photonic Bands of Metamaterials}
\titlerunning{Mie Scattering in the Macroscopic Response and the Photonic Bands of Metamaterials}
\author{Lucila Ju\'arez\textsuperscript{\Ast,\textsf{\bfseries 1}},
  Bernardo S. Mendoza\textsuperscript{\textsf{\bfseries 1}},
  and W. Luis Moch\'an\textsuperscript{\textsf{\bfseries 2}}}
\authorrunning{L. Ju\'arez, B. S. Mendoza and W. L. Moch\'an}
\mail{lucilajr@icf.unam.mx}
\institute{
  \textsuperscript{1}\,Centro de Investigaciones en \'Optica,
  Lomas del Bosque 115, Lomas del Campestre, 37150
  Le\'on, Guanajuato, M\'exico.\\
  \textsuperscript{2}\,Instituto de Ciencias F\'isicas, Universidad
  Nacional Aut\'onoma de M\'exico, Av. Universidad s/n, Col. Chamilpa, 62210
  Cuernavaca, Morelos, M\'exico}
\received{}
\published{}
\abstract{\bf%
We present a general approach for the numerical calculation of the
effective dielectric tensor of metamaterials and show that
our formalism can be used to study metamaterials beyond the long
wavelength limit. We consider a system
composed of high refractive index cylindric inclusions
and show that our method reproduces the Mie resonant features
and photonic band structure obtained from a multiple scattering approach,
hence opening the possibility to study arbitrarily complex geometries
for the design of resonance-based negative refractive metamaterials
at optical wavelengths. 
}

\begin{document}
\maketitle
\section{Introduction}

Metamaterials are usually composed of periodic arrangements of
micro- or nano- structures forming ordered patterns 
designed to control the propagation of light.
They can display exotic optical properties with interesting
aplications, such as a negative refractive index which
can be achieved exploiting underlying resonant features of the microstructure
\cite{shelby2001experimental,smith2004metamaterials,aydin2005investigation,hoffman2007negative,peng2007experimental}.
%
%
Resonant behavior has been vastly studied in split-ring-resonator (SRR)
structures, which are generally composed of metallic ring-like structures
which give rise to a significant magnetic response when excited by an
external inhomogeneous electric field.
However, metamaterials based on SRRs are not well suited to visible
frequencies, due to size limitations for its fabrication and high losses of the
metallic components
\cite{kuznetsov2016optically,SRR_THz,SRR_THz2}.
Recently, high refractive index inclusions have been investigated as
potential components for the fabrication of low-loss metamaterials displaying
strong magnetic properties at optical frequencies
\cite{kuznetsov2016optically,ZHAO200960,Kivshar17,ACSPhot_2017,PhysRevLett_102_133901,schuller2007dielectric,jahani2016all,zhang2018lighting}.
In dielectric materials, the displacement current increases with increasing
permittivity. In high refractive index particles the displacement current
can become large and give rise to the so-called Mie resonances at
wavelengths comparable with the size of the particles.
Thus, small high-index inclusions which display strong electric and magnetic
multipole Mie resonances in the optical region can be potentially used as
resonators for the fabrication of low loss negative refraction metamaterials.
Negative refraction has been reported for microstructure geometries as simple as
cylinders \cite{PhysRevLett_102_133901,Felbacq:94}.
Indeed, the number and frequencies of the resonant modes depend strongly
on the size and geometry of the particles.
Therefore, the generation and interference of such modes can be
controlled by manipulating the composition, size and shape
of the particles, allowing a wide range of possibilities
for the design of optical metamaterials.
However, general methods to compute the effective response of metamaterials
of arbitrary shape and composition are often limited to computationally expensive
numerical approaches or approximations. If the wavelength
of the incident light $\lambda$ is large compared to the
microstructure of the metamaterial, its response can be
described by an effective macroscopic dielectric function
efficiently computed within the
so-called {\it long wavelength limit}
\cite{mochan2010efficient,meza_second-harmonic_2019}.
Nevertheless, the description of Mie scattering lies by definition
outside the validity range 
of the long wavelength limit generally used to study metamaterials.
When the incident field varies in space on a length scale
comparable with the microscale of of the metamaterial, the effects of
retardation and non locality become important.
%
%
%
It has been shown that even when retardation effects are important, the system
may be characterized by a macroscopic dielectric response
\cite{P_rez_Huerta_2013}, which must be described by
a {\em non-local} tensor ${\bm \epsilon}({\bm r},{\bm r'},t,t')$
\cite{agranovich2006spatial}. 
This non-local retarded response results in a frequency $\omega$ and
wavevector ${\bm k}$ 
dependence of the effective dielectric tensor in Fourier space which leads to
the constitutive equation ${\bm D}(\omega, \bm k)={\bm \epsilon}(\omega,{\bm k})~{\bm E}(\omega,\bm k)$.
%

In this paper, we use a general efficient formalism for the numerical
computation of the effective dielectric response of
metamaterials at
arbitrary wavelenghts. We consider the case of a metamaterial
composed by high index dielectric cylinders and show that our approach
can reproduce 
the analytical results obtained from Mie theory at waveleghts comparable
with the size of the particle, thereby demonstrating that our method
can be used to study metamaterials based on Mie resonances.

The structure of the paper is the following. 
In Section \ref{Sec:Theory} we present our theory for the calculation
of the electromagnetic response of the metamaterial. First in Subsection
\ref{Subsec:Photonic_Ret} we develop an
efficient computational method based on the calculation of the macroscopic
dielectric response through a recursive procedure. This method is
applilcable to arbitrary materials and geometries. In order to
interpret and test its results, in Subsection
\ref{Subsec:Mie_theory} we develop a multiple scattering approach
applicable to an array of dielectric cylinders. 
Results on the
dielectric response of high-index 
dielectric cylinders are presented in Section \ref{Sec:Results},
compared to the results of the multiple scattering approach and
interpreted in terms of coupled Mie resonances. 
Finally, our conclusions are presented in
Section \ref{Sec:Conclusions}.

\section{Theoretical methods}\label{Sec:Theory}

\subsection{Macroscopic response}\label{Subsec:Photonic_Ret}

Recall that any physical vector field has a longitudinal part which
can be obtained 
by the projection operator ${\mathcal P}_L=\nabla\nabla^{-2}\nabla$ where
$\nabla^{-2}$ is the inverse of the Laplacian operator.
In reciprocal space this longitudinal projector can be written
as  $\hat{\mathcal P}_L ={\bm k}{\bm k}/{k^2}$,
where ${\bm k}$ is the wavevector of magnitude $k$.
%
Furthermore, for a periodic system with Bravais lattice
$\{ {\bm R} \} =\{ \sum^{D}_{i} n_{i}{\bm d}_{i} \}$ where $n_i$ are integers,
${\bm d}_{i}$ are primitive lattice vectors and $D$ is the number of dimensions,
the electric field can be expressed using Bloch's theorem as
\begin{equation}\label{Eq:E_Fourier_expansion}
  {\bm E_{\bm k}}({\bm r})= \sum_{\bm G}{\bm E_{G}} \exp({~i({\bm
      k+\bm G}) \cdot {\bm r}}),
\end{equation}
where ${\bm E_{\bm G}}$ is the amplitude of a plane wave with
wavevector ${\bm k+\bm G}$, with
${\bm G}$ a vector of the reciprocal lattice defined by
$\exp(~i{\bm G}\cdot{\bm R})=1$ and ${\bm k}$ a vector within
the first Brillouin zone. Here, the long wavelength limit corresponds
to $k\ll G$ \cite{mochan2010efficient}.
Notice that the field components with wavevectors ${\bm k+\bm G}$
fluctuate over distances 
of the order $d_i$, except for the term ${\bm G}=0$ which we identify as the
average of the field.
With this definition we can write an expression for the average operator as
\begin{equation}\label{Eq:Proj_op}
  \hat {\mathcal P}^{a}_{\bm G \bm G'}=\delta_{{\bm G}0}\delta_{{\bm G'}0}.
\end{equation}

Consider a binary metamaterial composed of a periodic lattice of
microstructures of arbitrary shape embedded in an homogeneous medium.
We define the structure function $\mathcal B$ which contains the
information on the shape of the inclusions
\begin{equation}\label{Eq:struc_func}
\mathcal{B}({\bm r}) = \left\{
\begin{array}{lr}
  1, & {\bm r} \in \mathcal{B},\\
  0, & {\bm r} \notin \mathcal{B}.
\end{array}
\right.
\end{equation}
For a system of two components, say, a
host ($A$) and inclusions ($B$), having each well defined dielectric functions
$\epsilon_A$ and $\epsilon_B$, the
{\it microscopic} dielectric function can be written as
\begin{equation}\label{Eq:eps_micro}
\epsilon({\bm r}) = \left\{
\begin{array}{lr}
  \epsilon_A, & {\bm r} \in A,\\
  \epsilon_B, & {\bm r} \in B,
\end{array}
\right.
\end{equation}
which we abbreviate in terms of the structure function as
$\epsilon({\bm r})=\frac{\epsilon_A}{u} \left(u -\mathcal{B}(r) \right)$,
where $u=1/(1-\frac{\epsilon_A}{\epsilon_B})$ is the spectral variable.

We begin by writing the wave equation for the electric field at
frequency $\omega$ using the free wavevector $q\equiv\omega/c$ as
\begin{equation}\label{Eq:wave_eq}
  \hat{\mathcal{W}}~{\bm E} = \left(\hat{\epsilon} + \frac{1}{q^2}
  \nabla^2 \hat{\mathcal{P}}_T \right) {\bm E} = 
   ~\frac{4\pi}{i \omega} ~{\bm J}_\mathrm{ext},
\end{equation}
which we solve formally as
\begin{equation}\label{Eq:wave_sol}
{\bm E} =\frac{4\pi }{i \omega} ~\hat{\mathcal{W}}^{-1}  ~{\bm
  J}_\mathrm{ext},   
\end{equation}
where we have introduced a wave operator $\hat{\mathcal W}$
written in terms of the transverse projector
$\hat{\mathcal P}_T =  \mathds{1} - \hat{\mathcal P}_L$.
Here, we identify the average of the field ${\bm E}^a=\hat{\mathcal P}^a{\bm E}$
in Eq.~(\ref{Eq:wave_sol}) with the {\it macroscopic} field ${\bm E}^M$.

Since ${\bm J}_\mathrm{ext}$ is an
external current, it does not have spatial fluctuations 
due to the microstructure and ${\bm J}^a_\mathrm{ext} = {\bm
  J}_\mathrm{ext}$. 
Thus, we average Eq. (\ref{Eq:wave_sol}) to obtain
\begin{equation}\label{Eq:wave_macro}
{\bm E}_M =\frac{4\pi }{i \omega} ~\hat{\mathcal{W}}_M^{-1}  ~{\bm
  J}_\mathrm{ext},   
\end{equation}
where the {\it macroscopic }wave operator $\hat{\mathcal{W}}_M$ is given by 
$\hat{\mathcal{W}}^{-1}_{M} = \hat{\mathcal{W}}^{-1}_{aa}=
\hat{\mathcal P}^a \hat{\mathcal{W}}^{-1}\hat{\mathcal P}^a$, i.e., its
inverse is the average of the inverse of the {\it microscopic} wave
operator.

Substitution of the permittivity tensor in terms of the structure function
Eq.~(\ref{Eq:struc_func}) and the spectral
variable $u$, leads to the wave-operator
\begin{equation}\label{Eq:wavevsB}
  \hat{\mathcal{W}} = \frac{\epsilon_A}{u} \left(u-\hat{\mathcal{B}}\right)
  + \frac{1}{q^2}\nabla^2 \hat{\mathcal{P}}_T,
\end{equation}
which we rewrite as
\begin{equation}\label{Eq:wavevsBg}
  \hat{\mathcal{W}} = \frac{\epsilon_A}{u} \left( u \hat{g}^{-1}
  -\hat{\mathcal{B}} \right)
\end{equation}
by introducing a {\em metric operator}
\begin{align}
  \hat{g} =&\left({\bm 1} + \hat{\mathcal{P}}_T \frac{\nabla^2}{q^2 \epsilon_A} \right)^{-1}.
\end{align}
Inverting the wave operator and taking the average we obtain
\begin{equation}\label{Eq:Waa^-1}
  \hat{\mathcal{W}}^{-1}_M =\hat{\mathcal{W}}^{-1}_{aa} =
    \frac{u}{\epsilon_A} \hat{g}_{aa} 
  \left(u-\hat{\mathcal{B}}\hat{g}\right)^{-1}_{aa},
\end{equation}
where we used the fact that the metric doesn't couple average to
fluctuating fields.

Finally, we extract the macroscopic dielectric tensor from the
corresponding wave operator
\begin{equation}\label{Eq:ret_diel_tens}
  \bm{\epsilon}^M (\omega, {\bm k}) = \frac{1}{q^2} (k^2 {\bm 1} - {\bm k}{\bm k}) +
  \bm{\mathcal{W}}^M (\omega, {\bm k}),
\end{equation}
where we used the explicit transverse projector for
a plane wave with wavevector ${\bm k}$.

In order to compute the macroscopic dielectric tensor, we begin
by calculating $(u - \mathcal{B}\hat{g})_{aa}^{-1}$ in Eq.~(\ref{Eq:Waa^-1}).
%
%
To that end, we first notice that the operator $\hat{\mathcal B}$ is
Hermitian in the usual sense. The operator $\hat g$ would also be
Hermitian if the response of medium A is dissipationless, i.e., if
$\epsilon_A$ is real. Nevertheless, the product $\hat{\mathcal B}\hat
g$ is not Hermitian. We notice, however, that the product $\mathcal{B}\hat{g}$
becomes Hermitian by
redefining the internal product between two {\em states} using  $\hat
g$ as a metric tensor. Thus, we define the $g$-product of two states
$\ket{\phi}$ and $\ket\psi$ as $(\phi|\psi)$, where
\begin{equation}\label{Eq:scalar}
  (\phi|\psi) \equiv \bra\phi\hat{g}\ket\psi,
\end{equation}
and $\braket{\ldots|\ldots}$ is the usual Hermitian scalar product. With this
definition it is clear that
\begin{equation}\label{Eq:herm}
  (\phi|(\hat{\mathcal{B}}\hat{g}|\psi)=
  \bra\phi \hat{g}\hat{\mathcal{B}}\hat{g}\ket\psi =
  \bra\psi \hat{g}\hat{\mathcal{B}}\hat{g}\ket\phi^* = 
  (\psi|\hat{\mathcal{B}}\hat{g}|\phi)^{*},
\end{equation}
so that $\hat {\mathcal B}\hat g$ is indeed Hermitian under the
product $(\ldots|\ldots)$ and we may borrow computational methods
developed for quantum mechanical calculations.

We choose an initial state $\ket{0} = b_0^{-1} \ket{p}$ where
$\ket{p}$ corresponds 
to a plane wave of frequency $\omega$, wavevector ${\bm k}$ and polarization $\hat{\bm e}$,
normalized as $\braket{p|p}=1$ under the conventional internal
product, and $b_0$ is 
chosen such that the state $\ket{0}$ is $g$-normalized,
$(0|0)=g_0=\pm1$. Notice that as $\hat g$ is not positive
definite, we should allow for negative {\em norms}.
We also define $\ket{-1}=0$.
Following the Haydock recursive scheme \cite{haydock1980recursive},
new states can be generated by repeatedly applying the
Hermitian operator 
\begin{equation}\label{Eq:haydock}
  \hat{\mathcal{B}}\hat{g} \left| n \right\rangle 
  = ~b_{n+1} \left| n+1 \right\rangle +  a_n \left| n \right\rangle +
  b_ng_ng_{n-1} \left| n-1 \right\rangle
\end{equation}
where the real coefficients $a_n$, $b_n$ and $g_n$ are obtained by imposing
the orthonormality condition
\begin{equation}
(n|m)=\braket{n|\hat g|m}= g_{n} \delta_{nm}
\end{equation}
and $g_n=\pm 1$.
Thus the operator $\hat{\mathcal{B}}\hat{g}$
has a tridiagonal representation in the basis ${\ket n}$,
which allows us to express the operator $(u - \hat{\mathcal{B}}\hat{g})$ as
%
%
\begin{multline}\label{Eq:ret_trid_matrix}
  \left( u - \hat{\mathcal{B}} \hat{g} \right) =\\\\
\begin{pmatrix}
  u-a_0 & -b_1 g_1 g_0 & 0 & 0 & \dots \\
  -b_1 & u-a_1 & -b_2 g_2 g_1 & 0 &  \\
  0 & -b_2 & u - a_2 & -b_3 g_3 g_2 &  \\
  0 & 0 & \ddots & \ddots & \\
  \vdots & & & & .\\
\end{pmatrix}
\end{multline}

Finally, we have to invert and average the operator in Eq.~(\ref{Eq:ret_trid_matrix}).
We recall that the average (Eq.~(\ref{Eq:Proj_op})) is given in terms
of a projection into our starting state $\ket{p}$, i.e., the zeroth row and
column element of the inverse operator which may be found for Eq.~(\ref{Eq:ret_trid_matrix})
in the form of a continued fraction
\begin{multline}\label{Eq:frac}
  \hat{\bm e} \cdot (\bm {\mathcal W}_M(\omega,\bm k))^{-1} \cdot \hat{\bm e} =\\\\ 
  \displaystyle{\frac{u}{\epsilon_A}} ~ \cfrac{g_0 b_0^2}{u-a_0 -
    \cfrac{g_0 g_1 b_1^2}{u-a_1-\cfrac{g_1 g_2 b_2^2}{u-a_2-
        \cfrac{g_2 g_3 b_3^2}{\ddots} }} }.
\end{multline}

Choosing different independent polarizations $\hat {\bm e}$ for the
initial state $\ket{p}$, one can compute all the independent
projections of the inverse 
of the wave tensor.
The result is then substituted in
Eq.~(\ref{Eq:ret_diel_tens}) 
to obtain the fully retarded macroscopic dielectric tensor. 
Further details on the method and its implementation can be found in
Refs.~\cite{P_rez_Huerta_2013,PSS_2018}. We remark
that the procedure above may be performed for two phase systems of
arbitrary geometry and composition as long as one of them is
dissipationless.

\subsection{Scattering approach}\label{Subsec:Mie_theory}

In this section we follow Ref.~\cite{Gagnon_2015} to compute the solution
of the multiple scattering problem of a finite array of dielectric cylinders.
Consider first a single infinitely long dielectric cylinder of radius
$R$ and refractive 
index $n$, standing in vacuum with its axis aligned to the $\hat{z}$ axis.
An incident field polarized on the $x-y$ plane is
applied. The magnetic field ${\bm H}=(0,0,\Phi)$ is taken parallel to the axis of
the cylinder, 
and satisfies scalar Helmholtz equations of the form
\begin{multline}\label{Eq:Helmholtz}
  \frac{1}{r}\frac{\partial}{\partial r}\left( r \frac{\partial
      \Phi^\beta(r,\theta)}{\partial r}\right)
    + \frac{1}{r^2}\frac{\partial^2
      \Phi^\beta(r,\theta)}{\partial^2 \theta} \\
    + \left( n^{\beta}(\omega) \frac{\omega}{c} \right)^2
    \Phi^\beta(r,\theta) = 0 
\end{multline}
for each frequency $\omega$, where $n^{\beta}=n$ or $1$ is the
refractive index of the region 
$\beta=I,O$, inside or outside of the cylinder, respectively.
Solutions of Eq.~(\ref{Eq:Helmholtz}) can be obtained as the products
$\Phi_l^\beta(\kappa^\beta)\exp(il\theta)$, where $l=0, \pm 1,
\pm 2, ...$ and $\Phi_l^\beta$ solves the Bessel
differential equation 
\begin{equation}\label{Eq:Bessel}
  \kappa^\beta\frac{d}{d\kappa^\beta}\left(\kappa^\beta \frac{d \Phi_l^\beta}{d
      \kappa^\beta}\right) + 
  \left( (\kappa^\beta)^2 - l^2 \right) \Phi_l^\beta =0,
\end{equation}
where $\kappa^{\beta}=n^{\beta} q r$.
The general solution of Eq.~(\ref{Eq:Bessel})
inside (I) and outside (O) the cylinder can be written in terms of the Bessel
functions of the first and second kind $J_l$ and $Y_l$ as
\begin{align}\label{Eq:Phi(r,theta)}
  \Phi^I(r,\theta) &= \sum_l  c_{l} J_l(nqr) \exp(il\theta),
  \\ \label{Eq:Phi(r,theta)_e}
        \Phi^0(r,\theta) &=  \sum_l
        \left[ a_{l}  J_l(qr) + b_{l} H_l(qr)\right] \exp(il\theta),
\end{align}
where we have chosen the outgoing Hankel functions $H_l=J_l+i Y_l$ as
the scattered field. The
coefficents $a_{l}$ 
describe the incident field and $b_{l}$ and $c_{l}$
are to be determined by the boundary conditions.
These are the continuity of $H$
and the continuity of the component of the electric field ${\bm E }=
(i/\epsilon q)\nabla \times {\bm H}$ parallel to the 
interface, $E_{\theta}$, at $r=R$. As
\begin{align}\label{Eq:E_IO}
        E^I_{\theta}(r,\theta) &= %
        \frac{i}{n}\sum_l c_l~ J_l'(nqr) \exp(il\theta)
        \\ \label{Eq:E_OO}
            E^0_{\theta}(r,\theta) &=
            i\sum_l  
            \left[ a_l J_l'(qr) + b_l  H_l'(qr)\right] \exp(il\theta)
\end{align}
where $J_l'$ and $H_l'$ denote the derivatives of the Bessel and
Hankel functions 
with respect to their arguments, then
\begin{equation}
  \begin{split}\label{Eq:BC}
  c_l J_l(nqR) &= a_l J_l(qR) + b_l H_l(qR), \\
  (1/n) ~c_l J_l'(nqR) &= a_l J_l'(qR) + b_l H_l'(qR),
  \end{split}
\end{equation}
which we solve for the scattering
coefficients
\begin{equation}\label{Eq:s_l}
  s_l \equiv \frac{b_l}{a_l} =
  \frac{J_l'(nqR)J_l(qR)-n~J_l'(qR)J_l(nqR)}{n~J_l(nqR)H_l'(qR)
    - H_l(qR)J_l'(nqR)}. 
\end{equation}

We consider now an array of $N \times N$ cilinders located at
positions $\{{\bm R}_n\}$. An 
incident plane wave traveling along the $x$-axis can be expressed in
the frame of reference of the $n_{th}$ cylinder as
\begin{equation} \label{Eq:incident}
  \Phi_{in}(\bm r) =  \exp (ik X_n) \sum_l i^{l} J_l(qr_n)\exp(il\theta_n),
\end{equation}
where $\bm R_n=(X_n,Y_n,Z_n)$ and $\bm r-\bm R_n$ is described by the
polar coordinates $r_n$ and $\theta_n$. 
The Graf's addition theorem allows us to rewrite a cylindrical function
centered at $\bm R_{n'}$ in a frame of reference centered at $\bm
R_n$. The wave 
scattered by cylinder ${n'}$ can be rewritten in the frame
of reference of cylinder $n$ as
\begin{multline}\label{Eq:Graf}
  H_{l'}(qr_{n'}) \exp (il'\theta_{n'}) = 
   \sum_l \exp (i(l-l')\phi_{nn'}) \times \\ H_{l-l'}(qR_{nn'}) J_l(qr_n) \exp(il\theta_n)
\end{multline}
where $\bm R_{nn'}=\bm R_n-\bm R_{n'}$ is described by the polar
coordinates $R_{nn^{'}}$ and $\phi_{nn^{'}}$.
Thus, the magnetic field in the interstices may be described in
coordinates centered at the  $n_{th}$ cylinder as the sum of the
incident field, the $n$-th scattered field and the field scattered by
all the other cylinders, 
\begin{multline}
  \label{Eq:Phi_outside}
  \Phi^O(\bm r) = 
  \exp (ikX_n) \sum_l i^{l} J_l(qr_n)\exp(il\theta_n)\\
  +  \sum_l b_{nl} H_l(qr_n) \exp(il\theta_n)\\ 
  + \sum_{n'\ne n}\sum_{ll'}
  b_{n'l'}\exp (i(l-l')\phi_{nn'}) \\
   \times H_{l-l'}(qR_{nn'}) J_l(qr_n) \exp(il\theta_n)
\end{multline}
From Eqs.~(\ref{Eq:Phi(r,theta)_e}) and (\ref{Eq:Phi_outside}) we
identify the coefficients $a_{nl}$ as 
\begin{multline}
  \label{Eq:a_nl}
  a_{nl}=  \exp (ikX_n) i^{l} +
  \sum_{n'\neq n}\sum_{l'}
  \exp (i(l-l')\phi_{nn'}) \\
   \times H_{l-l'}(qR_{nn'}) b_{n'l'}.
\end{multline}
Introducing the scattering coefficients $s_{nl}$  of each
cylinder as in (\ref{Eq:s_l}) yields
\begin{multline}
  \label{Eq:Tb_nl=a_0l}
  b_{nl} - s_{nl} \sum_{n'\neq n} \sum_{l'}
  \exp (i(l-l')\phi_{nn'})  \\ \times H_{l-l'}(qR_{nn'}) b_{n'l'} 
  =  s_{nl} \exp (ikX_n) i^{l},
\end{multline}
which can be summarized into a system of coupled equations
\begin{equation}
  \label{Eq:Tb=a}
  \bm T~\bm b=\bm a,
\end{equation}
where ${\bm a}=\{ s_{nl}\exp (ikX_n) i^{l}\}$ describes the incident
field, ${\bm b}=\{ b_{nl}\}$ describes the scattered field in the
interstitial region, to be obtained, and
\begin{multline}
  \label{Eq:T}
  T^{ll'}_{nn'} = \delta_{nn'}\delta_{ll'}  - (1-\delta_{nn'}) \times \\
  \exp (i(l-l')\phi_{nn'}) H_{l-l'}(qR_{nn'})s_{nl}.
\end{multline}

\section{Results}\label{Sec:Results}

We consider a metamaterial made of a square lattice of
identical infinitely long dielectric cylinders of
radius $R$ and refractive index $n$ set in vacuum,
with a lattice constant $a$.
Using an efficient computational implementation \cite{Photonic} of the
numerical approach 
presented in Sec.~\ref{Subsec:Photonic_Ret}, we calculate
the macroscopic dielectric function of the metamaterial as a function
of frecuency $\omega$ and wavevector $k$. We compare our results with
the analytical solution obtained in Sec.~\ref{Subsec:Mie_theory}
for a finite array of $N \times N$ cylinders.

We first consider thin weakly
interacting cylinders with radius $R=0.1a$ and refractive index $n=10$.
Numerical calculations were performed using a 2D  $601\times601$ 
grid to discretize the unit cell and
performed the recursive calculation using $450$ Haydock coefficients.
For the analytical case we considered a large finite array of
$511\times511$ cylinders 
and a maximum value of the orbital number $l=1$.
We checked the convergence of the results by repeating the
calculations with larger grids, more Haydock coefficients and larger
angular momenta.
\begin{figure}[t]
  \begin{center}
    \includegraphics[scale=0.75]{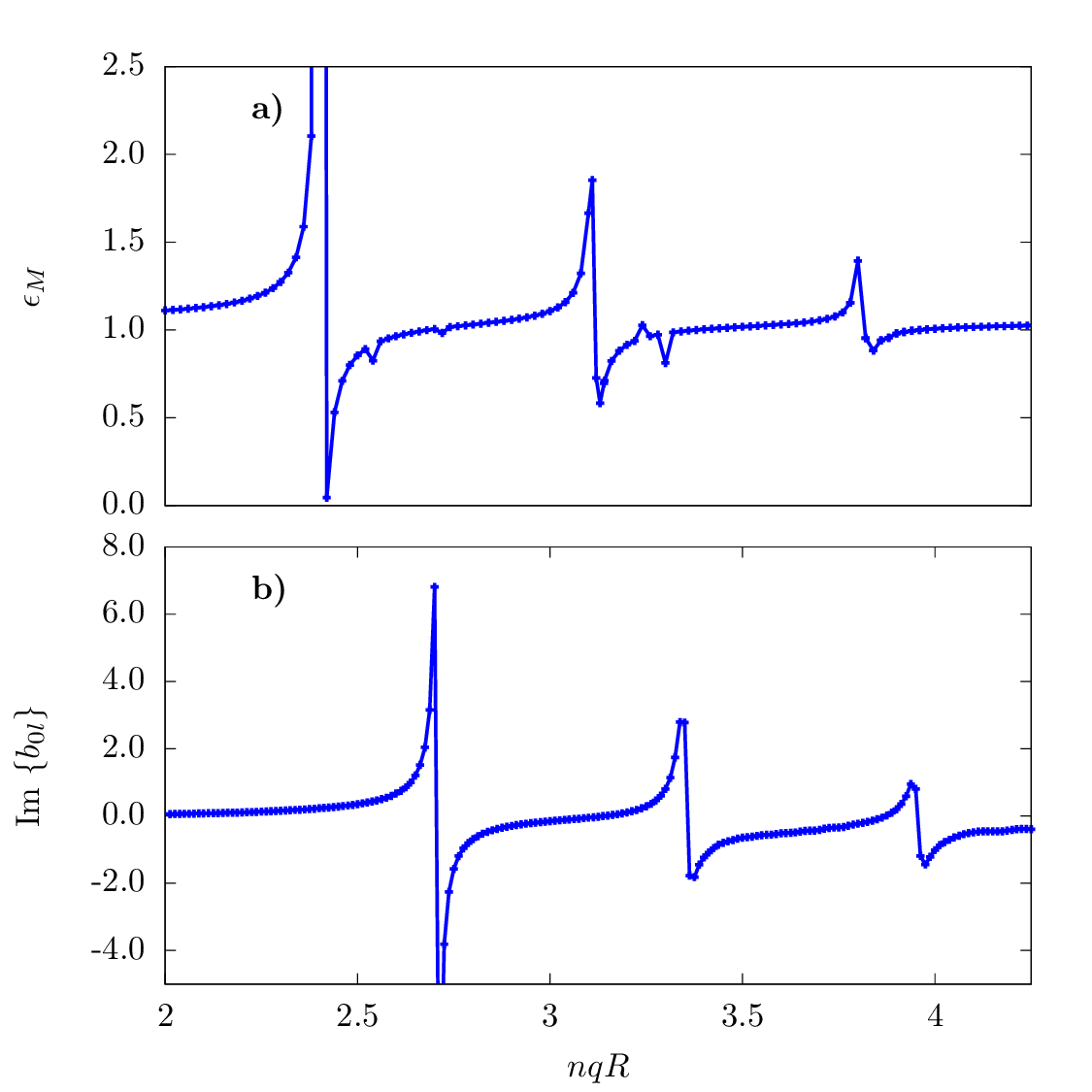}
    \caption{\label{Phot_vs_Mult} (a) Transverse component
      $\epsilon_M^T=\epsilon_M^{yy}(\omega,\bm k)$ of the macroscopic
      dielectric response of a metamaterial composed
      of a square lattice of dielectric cylinders of refraction index $n=10$ and
      radius $R=0.1a$ as a function of the frequency, characterized by
      $nqR$, for a wavevector along the $x$ direction slightly larger
      than the vacuum wavevector $k=1.01q$. (b) Sum of the imaginary
      part of coefficients of 
      the scattered field $\sum_lb''_{0l}$ of an array of $511 \times
      511$ dielectic cylinders, obtained from Eq.~(\ref{Eq:Tb=a})
      using a maximum value of $l=1$.} 
  \end{center}
\end{figure}
Fig. \ref{Phot_vs_Mult}a shows the results for the transverse
component of the macroscopic dielectric tensor of the metamaterial
$\epsilon_M^T=\epsilon_M^{yy}$ 
obtained through the recursive numerical approach.
The results are shown as a function of the frequency, characterized by
the free wavevector within the dielectric normalized to the radius $nqR$.
Several resonances are clearly visible. 
In order to analyze their origin, in Fig. \ref{Phot_vs_Mult}b
we show the sum of the imaginary part
of the scattered field coefficients $\sum_lb''_{0l}$ obtained from the
analytical method. 
In this calculation we assumed the response of all cylinders was
identical, except for the phase factor $\exp(ikX_n)$ and we included a
damping of the cylinder-cylinder interaction at large distances to
eliminate the oscillations due to reflections at the edge of the
finite array, and thus mimic an infinite array. We checked convergence
of this procedure by increasing the number of cylinders.

Three prominent resonant features are observed in both panels of
Fig. \ref{Phot_vs_Mult} at low energies.
The lowest energy resonance corresponds to a magnetic dipole arising
from the term $l=0$, as it lies close to that of an isolated
cylinder ocurring at the first zero of the Bessel function
around $nqR \approx 2.4$.
The second resonance  around $nqR \approx \pi$, emerges from the
fulfillment of Bragg's diffraction condition for $2a=2\pi/q$.
A third resonance close to $nqR \approx 3.8$ is originated by the term $l=1$.
This resonance is strongly enhanced through the interaction between cylinders.
The additional peaks in the macroscopic response are due to resonances
caused by multiple reflections in the intesrstitial regions. We have
verified that they are not due to numerical noise, but they dissapear
when a very small artificial dissipation is 
added to the interstitial dielectric function, while the large peaks
are robust. 
Notice however that the scattering coefficients $b_{0l}$ resonate at
higher energies than the macroscopic dielectric function. The reason
for this discrepancy is that the transverse normal modes of the system
are not actually given by the poles of the dielectric response, but by
the poles of the electromagnetic Green's function.

\begin{figure}[b]
  \begin{center}
    \includegraphics[scale=0.75]{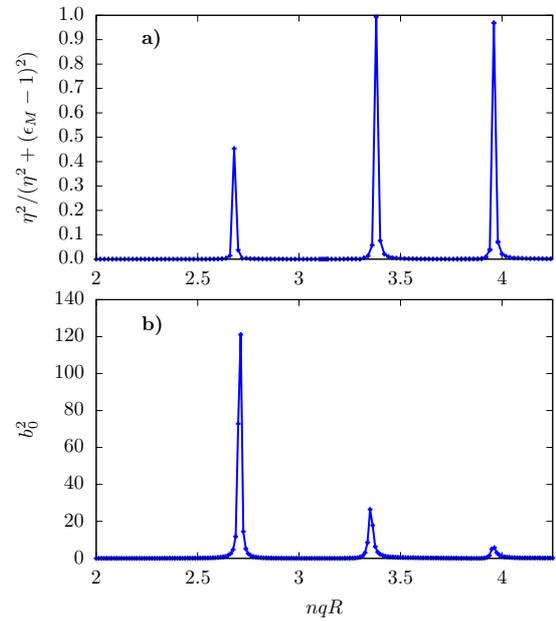}
    \caption{\label{Green_Mie}a) Normalized imaginary
      part of the electromagnetic 
      Green's function $\eta^2/(\eta^2+(\epsilon_T -k^2/q^2)^2)$ for a
      small dissipation parameter $\eta$
      as a function of $nqR$ for $k\approx q$ for
      the same system as in Fig. \ref{Phot_vs_Mult}.
      b) Squared magnitude of the scattering amplitude
      $b_0=\sum_lb_{0l}$ for the same system 
      as in Fig. \ref{Phot_vs_Mult}.}
  \end{center}
\end{figure}

Figure \ref{Green_Mie} a) shows the imaginary part of the
electromagnetic Green's function $(\epsilon_T -k^2/q^2)^{-1}$
for a small broadening parameter
$\eta=0.001$ and b) the squared magnitude of the scattering amplitude
$b_0=\sum_l b_{0l}$ for $k\approx q$ as a function of $nqR$.
Here, the peaks of the Green's function coincide with the peaks of the
scattering coefficient as the poles of the
Green's function and of the scattered coefficient correspond both to the
normal modes of the system, for which one may have a finite field, and
finite scattered amplitudes without an external excitation.

Having verified the consistency of both computational approaches when
applied to the calculation of the normal modes of a system,  
we now consider a more interesting and realistic case. We consider a
metamaterial composed of strongly interacting cylinders with a larger radius
$R=0.35a$ and a large but realistic refraction index 
$n=4$ (similar to that of Si).
In order to identify exotic behavior such as negative refraction, we
have to examine the dispersion relation of the normal modes of the
system and explore their group velocity
\cite{agranovich2006spatial,agranovich2009electrodynamics}.
Thus, we calculated the macroscopic dielectric function and obtained 
the Green's function of the metamaterial as a
function of both frecuency $\omega$ and wavevector $k$.
\begin{figure}[t]
  \begin{center}
    \includegraphics[scale=0.7]{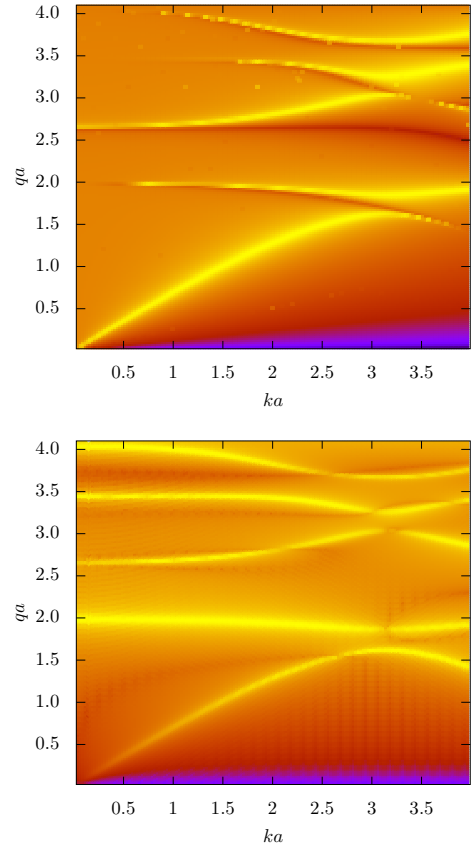}
    \caption{\label{disp_rel} Imaginary part of 
      the Green's function $(\epsilon_T -k^2/q^2)^{-1}$ for
      a system composed by dielectic cylinders of radius $r=0.35a$ and
      refractive index $n=4$, 
      obtained numerically through the macroscopic response using the
      package {\em Photonic} (upper panel).
      Magnitude of the scattered amplitude $b_0$ normalized as
      $\eta/(\eta^2 + 1/b_0^2)$ using a dissipation factor $\eta$
      for an array of $201 \times 201$ cylinders, obtained through the
      scattering approach (Eq.~(\ref{Eq:Tb=a})) 
      using a maximum value of $l=2$ (lower panel).
      Color map is logarithmic.}
  \end{center}
\end{figure}
For the numerical calculations we used a two dimensional
$201\times201$ grid and $300$ sets of Haydock coefficients.
A very small artificial dissipative term $0.001i$ has been added to the
vaccum dielectric constant in order to improve convergence. The
results are displayed in Fig.~\ref{disp_rel}.
The scattering amplitude obtained from the scattering
approach, calculated for an array of $201\times201$ cylinders
considering a maximum value of the orbital number $l=2$ is shown
for comparisson.

The upper panel of Fig.~\ref{disp_rel} shows the imaginary part of the
of the Green's function $\eta/(\eta^2 +(\epsilon_M^T -k^2/q^2))$ where
$\epsilon_M^T$ was obtained numerically through Haydock's recursion
and the results have been smoothed with a dissipation factor $\eta=0.2$
for better visualization. 
The lower panel shows the absolute value squared of the scattering
amplitude $b_0=\sum_l b_{0l}$, smoothed as
$\eta/(\eta^2 + 1/|b_0|^2)$ with a dissipation factor $\eta=0.1$.
\begin{figure}[b]
  \begin{center}
    \includegraphics[scale=0.7]{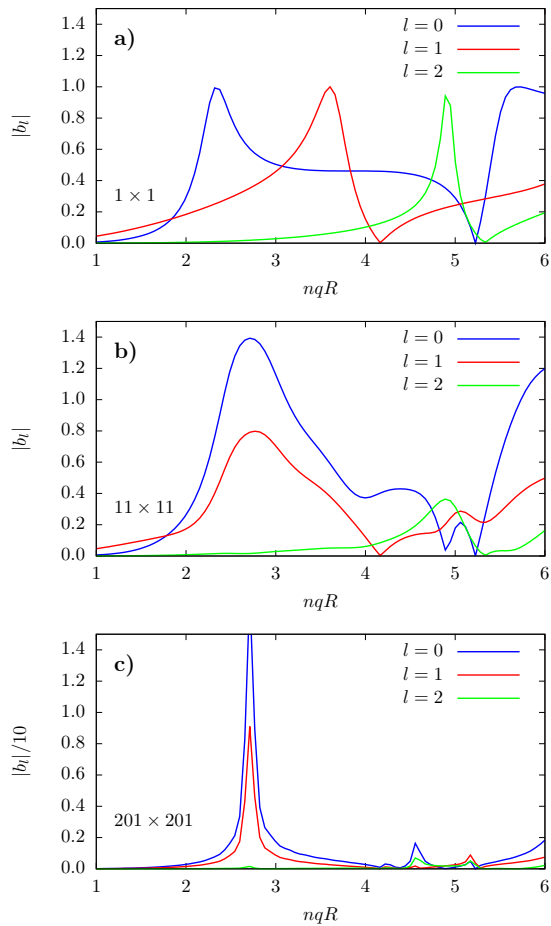}
    \caption{\label{bands} Absolute value of the scattering
      coefficients $|b_{0l}|$ 
      for an a) isolated cylinder, b) array of $11\times11$ cylinders and c)
    array of $201\times201$ cylinders, as a function of $nqR$.}
  \end{center}
\end{figure}

Several bands may be identified in Fig. \ref{disp_rel} showing a
very good agreement between the dispersion relations obtained through
our two approaches. 
Regions of negative dispersion, i.e., for which the frequency
of the resonances decreases as the wavevector increases, yielding a negative
group velocity, are clearly observed for the second, fourth and fifth
bands and for large wavevectors before the first Brillouin zone
boundary at $ka=\pi$.
The bands originate from the combination of isolated cylinder resonances
of different values of $l$, while different values of $l$ participate in
each band due to the relatively strong interaction between neighboring
thick cylinders.
This can be confirmed through Figure~\ref{bands}, which shows the
absolute value of the individual scattering coefficients $b_{0l}$ for
$l=0,1,2$ as a function of $nqR$ for 
a) a single cylinder, b) an array of $11\times11$ cylinders and
c) an array of $201\times201$, for a wavevector $k=q$.
The coefficients corresponding to an isolated cylinder display one
broad peak each 
around $q=2.4$, $4.8$ and $5.1$ for $l=0,1,2$ respectively, which are
close to the first zeroes of the corresponding Bessel function.
For an array of $11\times11$ cylinders these peaks appear distorted
and shifted to be finally merged for a larger array of
$201\times201$ in which case all $l$-contributions resonate
at the same frequencies.

\section{Conclusions}
\label{Sec:Conclusions}
We presented two schemes to calculate the electromagnetic
properties of a metamaterial made of a simple lattice of
cylinders: a numerical method based on a recursive calculation of the
macroscopic dielectric tensor which may be easily generalizable to
arbitrary geometries and materials, and a
multiple scattering approach for cylindrical geometries
which allowed us a simple interpretation of the results in terms of the
excitation of Mie resonances. We applied these methods to investigate 
the response of a system made up of cylinders of high
index of refraction. The comparison between the results of
both methods is not direct, as in one case we obtain a macroscopic
response and in the other we obtain scattering
amplitudes. Nevertheless, we showed that the poles of the macroscopic
Green's function obtained by our numerical method coincide with those
of the scattering coefficients, and both can be interpreted as the excitation of the
normal modes of the system.
For a system of thin cylinders with very high index of
refraction and a relatively small coupling, we identified the
nature of each mode. We found modes arising from the Mie
resonances of individual cylinders and a mode arising from Bragg
coherent multiple scattering.
For larger cylinders the interaction
yields a coupling between resonances with different angular
momenta. By varying the frequency and wavevector
independently we computed the dispersion relation of the normal
modes. The photonic band structure obtained using both methods is 
in very good agreement, and reveal regions of negative
dispersion.
Thus, through comparison with an ad-hoc model we showed
that our macroscopic approach based on Haydock's recursion and its
implementation in the {\em Photonic} package is an
efficient procedure for obtaining the optical properties of
metamaterials made up of high index of refraction
cylinders incorporating resonances which cannot be explored within the
long wavelength limit. Furthermore, as it can be readily generalized
to arbitrary geometries and materials, we believe it will prove to be
a useful tool 
for the design of artificial materials with a richer geometry that
might yield novel properties.

\begin{acknowledgement}
  This work has been supported by CONACyT through a postdoctoral
  research fellowship. WLM acknowledges the support from DGAPA-UNAM
  through grant IN111119.
  BSM acknowledges the support from CONACYT through grant A1-S-9410.
\end{acknowledgement}

\bibliographystyle{pss}
\bibliography{biblio}

\end{document}